\documentstyle[aaspp4,12pt]{article}
\begin{document}

\title{DOES THE RADIATIVE AVALANCHE FUELING WORK IN ANY ACTIVE GALACTIC NUCLEI ?}
\author{Yoshiaki Taniguchi}

\vspace {1cm}

\affil{Astronomical Institute, Tohoku University, Aoba, Sendai 980-77, Japan}


\begin{abstract}

Recently Umemura, Fukue, \& Mineshige (1997) proposed 
the radiative avalanche fueling to active galactic nuclei;
gas accretion is driven by radiation drag
exerted by stellar radiation from circumnuclear starburst regions.
This mechanism is also interesting in terms of starburst-AGN connections.
We therefore present observational tests for the radiative avalanche fueling.
Our tests, however, show that gas accretion rates driven by the radiative avalanche 
are significantly
lower than those expected from the standard accretion theory applied for
typical active galactic nuclei with the circumnuclear starburst regions. 
Instead we propose an alternative, possible starburst-AGN connection;
a minor merger with a nucleated satellite drives circumnuclear starbursts and 
then leads to gas fueling
onto the central engine as the merger proceeds.

\end{abstract}




\vspace{1cm}

\begin{center}

Accepted for publications in {\bf THE ASTROPHYSICAL JOURNAL (LETTERS)}

{\it 1997 July}

\end{center}

\section{INTRODUCTION}

Since the starburst activity is observed in the circumnuclear regions
of some active galactic nuclei (AGNs), 
possible starburst-AGN connections have been discussed for these two decades
(Weedman 1983; Terlevich \& Melnick 1985; Norman \& Scoville 1988; 
Sanders et al. 1988; Heckman et al. 1989; Rieke 1992;
Scoville 1992; Taniguchi 1987, 1992; Mouri \& Taniguchi 1992; 
Terlevich et al. 1992). 
The discussion has been made in  terms of; a) formation of compact 
nuclei (e.g., Weedman 1983), b) AGN modeling  without a supermassive black hole
(e.g., Terlevich et al. 1992), or c) gas fueling triggered by the circumnuclear
starbursts (e.g., Norman \& Scoville 1988; Taniguchi 1992).

Recently, Umemura, Fukue, \& Mineshige (1997a, 1997b; hereafter UFM97)
proposed a novel idea that 
mass accretion onto a supermassive black hole may be triggered by
radiation drag exerted by stellar radiation from the circumnuclear 
starburst regions (the radiative avalanche fueling),
based on the well-known physical process, so-called
Poynting-Robertson effect. 
Since it is known that a number of AGNs have actually luminous circumnuclear 
starburst (hereafter CNSB) regions (e.g., Telesco et al. 1984; 
Heckman et al. 1986; Keel 1987; 
Wilson 1988; Storchi-Bergmann et al. 1996a, 1996b
and references therein), 
it is worth testing if this new mechanism works from 
an observational point of view.

\section{A BRIEF SUMMARY OF THE RADIATIVE AVALANCHE FUELING}

We give a brief summary of the radiative avalanche fueling proposed by UFM97.
The radiative avalanche means that gas accretion is driven by radiation drag
exerted by stellar radiation from circumnuclear starburst regions.
If a surface layer of a rotating nuclear gas disk is irradiated by the
massive stars in the CNSB regions, the gas on the surface could lose
angular momentum via radiation drag
(i.e., Poynting-Robertson effect; Poynting 1903; Robertson 1937),
giving rise to an avalanche of the gas layer.

If a CNSB region with a total luminosity of $L_{\rm CNSB}$
is located at a radius $R$ with an effective size
$\Delta R$ (i.e., the effective diameter of a star-forming clump),
the accretion rate due to the avalanche at a radius $r$ ($< R$) is given by

\begin{equation}
{\dot M}_{\rm rad}  \simeq 0.2 ~ \eta_{\rm r} ~  {\rm sin} \theta_d ~
\left({r \over R} \right)^2 
\left({L_{\rm CNSB} \over 3\times 10^{12} ~ L_\odot} \right) ~~ M_\odot ~ {\rm y}^{-1}
\end{equation}

\noindent where $\eta_{\rm r}$ is the efficiency of 
radiative accretion ($\sim 0.1$ - 1), and 
(sin $\theta_d) \times (r/R)^2$ is the irradiation efficiency onto
the surface layer at radius $r$ where  
sin $\theta_d ~ \sim  \Delta R / (R - r)$.
In the following discussion, we adopt $\eta_{\rm r} = 1$.
In UFM97, the factor of sin $\theta_d$ is not seriously taken into account.
However, the recent {\it Hubble Space Telescope} observations have shown that
the CNSB regions consist of a number of star-forming clumps and each clump
(i.e., a star cluster) has an effective diameter of $\sim$ 10 pc at most
(Barth et al. 1995; Maoz et al. 1996).
This compact nature is also observed in the archetypical giant HII
region, 30 Dor in LMC (Weigelt et al. 1991).
Given the typical radius of 1 kpc for the CNSB rings 
(Storchi-Bergman et al. 1996a, 1996b), we estimate sin $\theta_d \sim 0.01$.
Furthermore,  the geometrical dilution factor is estimated to be
$(r/R)^2 \sim 0.01$ for $r$ = 100 pc and $R$ = 1 kpc.
Therefore, the accretion rate driven by the radiative avalanche is 
much lower than that estimated in UFM97.

The characteristic accretion timescale for the radiative avalanche
is estimated as

\begin{equation}
t_{\rm rad}  \simeq 2.4\times 10^6
\left({L_{\rm CNSB} \over 3\times 10^{12} ~ L_\odot} \right)^{-1}
\left({R  \over 100 ~ {\rm pc}} \right)^{2}
\left({f_{\rm dg}  \over 10^{-2}} \right)^{-1}
\left({a_{\rm d}  \over 0.1 \mu{\rm m}} \right)
\left({\rho_{\rm s}  \over  {\rm g ~ cm}^{-3}} \right)  ~~{\rm yr}
\end{equation}

\noindent where 
$f_{\rm dg}$ is the dust-to-gas  mass ratio, $a_{\rm d}$ is the size
of dust grains, and $\rho_{\rm s}$ is the mass density of solid material 
within the grain.
The CNSBs in AGNs are usually observed as {\it ringed} (or armed) star-forming clumps
and their typical radii range from several 100 pc to 1 kpc
(Telesco et al. 1984; Boer \& Schulz 1993;  Genzel et al. 1995;
Storchi-Bergmann, Wilson, \& Baldwin 1996; Storchi-Bergmann et al. 1996b;
Maoz et al. 1996).
The luminosities of the CNSBs amount to  $\sim 10^{11} L_\odot$ in
both NGC 1068 (e.g., Telesco et al. 1984) and NGC 7469 (e.g., Genzel et al.
1995).
If we adopt both $L_{\rm CNSB} = 10^{11} L_\odot$ and $R$ = 1 kpc, 
we obtain $t_{\rm rad} \sim 7 \times 10^9$ years.
This timescale seems  too long to cause efficient fueling to AGN
although it is known that the ages of the CNSBs
in Seyfert galaxies are estimated to be $\sim 10^8$ years,
being older significantly than those of on-going starbursts,
$\sim 10^7$ years (Glass \& Moorwood 1985;
Mouri \& Taniguchi 1992; Taniguchi \& Mouri 1992;
Dultzin-Hacyan \& Benitez 1994;  Oliva et al. 1995; Hunt et al. 1997).

\section{OBSERVATIONAL TESTS FOR THE RADIATIVE AVALANCHE FUELING}

\subsection{\it Circumnuclear Starbursts around Seyfert Nuclei}

Prior to the observational tests for the radiative avalanche fueling,
we argue the occurrence of the CNSBs around Seyfert nuclei.
Based on the {\it IRAS} data base, it has been discussed that Seyfert 2 nuclei
(hereafter S2s) tend to show excess 
mid- (MIR) and far-infrared (FIR) emission with resect to Seyfert 1s (S1s)
(Rodr\'iguez-Espinosa, Rudy, \& Jones 1987; Dahari \& DeRobertis 1988; Heckman et al. 1989; 
Maiolino et al. 1995). Also, molecular gas content is higher in S2s than in S1s (Heckman
et al. 1989). These observations are usually interpreted as that S2s tend to have
luminous CNSB regions more frequently  than S1s.
However, if there would be some extended dusty clouds around 
Seyfert nuclei, they show excess emission
at FIR because the equilibrium temperature is as cold as 100 K
(Pier \& Krolik 1993; Granato, Danese, \& Francheschini 1996; Taniguchi et al. 1997).
This means that S2s may have brighter FIR emission if they would  have more
extended dusty regions around their nuclei.
Thus we cannot  conclude that almost all S2s have the CNSB
{\it solely} from the analysis of FIR data.

Pogge (1989) made a narrow-band emission-line imaging survey of 20 nearby
Seyfert galaxies and found that the CNSB regions are present
only in $\sim$ 30 percent of the S2s while no CNSB is found in the S1s.
Although there is a famous S1 with the intense CNSB, NGC 7469 (Heckman et al. 1986;
Wilson et al. 1986, 1991; Keto et al. 1992; Mauder et al. 1994; Genzel et al. 1995),
it is known that there is a few S1s with CNSBs (Oliva et al. 1995; Hunt et al. 1997).
Hunt et al. (1997) showed form their NIR multi-color imaging study that
there is little evidence for CNSBs in S1s and even if CNSB events would occur in them,
they occurred more than $10^9$ years ago. 
Thus we consider that the radiative avalanche triggered by the CNSBs cannot 
work in most S1s in the nearby universe because of the absence of CNSBs in them.
Therefore, possible candidates of 
Seyfert nuclei triggered by the radiative avalanche may be
about one third of S2s and a few S1s which have the luminous CNSBs around the nuclei.
In the following subsections, we present observational tests
for typical Seyfert and LINER\footnote{LINER = Low Ionization Nuclear
Emission-line Region (Heckman 1980).} nuclei with CNSBs. Further we  mention about
the case of ultraluminous infrared galaxies (Sanders et al. 1988).

\subsection{\it Seyfert Nuclei with Luminous Circumnuclear Starbursts}

Among the Seyfert nuclei with CNSBs (Storchi-Bergmann et al. 1996a, 1996b and 
references therein), we study the two archetypical Seyfert nuclei, NGC 1068 (S2) and
NGC 7469 (S1). In order to examine if gas accretion driven by the radiative avalanche is 
high enough to achieve the observed bolometric luminosities
of central engines, we compare the two
accretion rates; 1) the gas accretion rate driven by radiative avalanche
due to the stellar lights form the CNSBs
[see equation (1)], and 2) the gas accretion rate estimated
with the standard accretion theory (Rees 1984),

\begin{equation}
{\dot M}_{\rm acc} = {L_{\rm bol} \over \eta_{\rm acc} c^2}  \simeq 0.02 ~ 
\left({L_{\rm bol} \over 10^{44} ~ {\rm erg~ s}^{-1}} \right)
\left({\eta_{\rm acc} \over 0.1} \right)^{-1}
 ~~ M_\odot ~ {\rm y}^{-1}
\end{equation}

\noindent where $\eta_{\rm acc}$ is the conversion efficiency from
the gravitational energy to the radiation. As for AGNs with CNSBs,
UFM97 considered that only high-energy photons come from the central engine
while the other radiation (e.g., from FIR to UV) comes form the CNSBs.  
Thus they used the X-ray luminosity instead $L_{\rm bol}$ when they estimated
${\dot M}_{\rm acc}$. However, this assumption seems inadequate.
The reason for this is, for example, 
that much amount of infrared emission of such galaxies comes from
the dusty tori (Pier \& Krolik 1993; Pier et al. 1994; see for the extreme case,
Taniguchi et al. 1997). Therefore, in the later discussion, 
we use the $L_{\rm bol}$ derived from
the spatially-resolved observations in the  estimates of ${\dot M}_{\rm acc}$. 
Also, we use the CNSB luminosities 
based on the spatially-resolved observations in the estimates of ${\dot M}_{\rm rad}$.
Our method makes it possible to perform accurate tests.

NGC 1068 is one of the most luminous Seyfert galaxies in the local universe.
Its bolometric luminosity of the central engine is estimated to be $L_{\rm bol} \simeq
8.5 \times 10^{44}~ (f_{\rm refl}/0.01)^{-1} (D/22 {\rm Mpc})^2$ erg s$^{-1}$ where
$f_{\rm refl}$ is the fraction of nuclear flux reflected into our line of sight
and $D$ is the distance to NGC 1068 (Pier et al. 1994). Adopting  the fiducial values
in Pier et al. (1994), we obtain the gas accretion rate,
${\dot M}_{\rm acc} \simeq 0.17 ~ M_\odot$ y$^{-1}$.
Next we estimate the accretion rate driven by radiative avalanche.
The bolometric luminosity of the CNSB regions is estimated to be
$L_{\rm CNSB} \simeq 2.2 \times 10^{11} L_\odot$ at distance of 22 Mpc
(Telesco et al. 1984). The radius of the CNSB ring is about 1 kpc
(Telesco et al. 1984; Baldwin et al. 1987)\footnote{Braatz et al. (1993) 
and Cameron et al. (1993) detected
the inner MIR component with size of $\sim$ 100 pc. Since, however,
the dusty torus in NGC 1068 is considered to be extended at radius of 100 pc
(Pier \& Krolik 1993), this MIR component may not be star-forming regions.
Even if this component is the star-forming regions, the accretion rate
by radiative avalanche is still smaller by one order than the predicted value.}.
Although there is no direct measurement of the vertical width ($\Delta R$)
 of the star-forming
clumps, it is estimated to be 10 pc at most (Weigelt et al. 1991; 
Barth et al. 1995; Maoz et al. 1996). If we consider the radiative avalanche
at $r$ = 100 pc, the irradiation efficiency is
sin$\theta_d ~ (r/R)^2 \simeq \Delta R~ r^2 / [(R - r) R^2] \sim 1.1 \times 10^{-4}$. 
We thus obtain the accretion rate by radiative avalanche,
${\dot M}_{\rm rad} \simeq 1.6 \times 10^{-6}~ M_\odot$ y$^{-1}$
for $r = 100$ pc.
Since this accretion rate is much smaller than that estimated from the accretion theory,
we consider that the radiative avalanche does not work in NGC 1068.

NGC 7469 is also one of the famous Seyfert galaxies with the luminous CNSB regions
(Cutri et al. 1984; Heckman et al. 1986; Wilson et al. 1986, 1991; 
Miles, Houck, \& Hayward 1994; Mauder et al. 1994; Genzel et al. 1995).
The bolometric luminosity of the central engine including the stellar 
luminosity of the host galaxy
is  $8.9\times 10^{44}$ erg s$^{-1}$
at a distance of 98 Mpc (Genzel et al. 1995). Since the contribution of the central 
engine to this luminosity 
is $\sim$ 40 percent (Kotilainen et al. 1992), we obtain 
$L_{\rm bol} \simeq 3.5 \times 10^{44}$ erg s$^{-1}$ and thus 
the required  gas accretion rate
is ${\dot M}_{\rm acc} \simeq 0.07 ~ M_\odot$ y$^{-1}$.
On the other hand, the CNSB ring has the luminosity of $4.5 \times 10^{11} L_\odot$
at a radius of 720 pc. Assuming $\Delta R$ = 10 pc again, we obtain 
${\dot M}_{\rm rad} \simeq 9.3 \times 10^{-6}~ M_\odot$ y$^{-1}$
for $r = 100$ pc,
being much lower than the expected value.
Accordingly, we have shown that the contribution of 
the gas accretion driven by radiative avalanche is negligibly small
in both the Seyfert nuclei, NGC 1068 and NGC 7469.

\subsection{\it LINERs with Circumnuclear Starbursts}

It is also known that some LINERs
have CNSBs (Storchi-Bergmann et al. 1996a, 1996b
and references therein). Here we investigate a case of NGC 1097
(Keel 1983; Hummel, van der Hulst, \& Keel 1987; 
Storchi-Bergmann, Baldwin, \& Wilson 1993; Storchi-Bergmann
et al. 1995; Barth  et al. 1995).
Although this galaxy was originally classified as a LINER (Keel 1983), 
the double-peaked  broad line emission appeared since 1991 (Storchi-Bergmann et al.
1993, 1995). In terms of the standard model, some sporadic accretion events would
occur since 1991 (Eracleous et al. 1995).
It is interesting to examine if this accretion is due to the radiative avalanche.
Although there is no measurement of the bolometric luminosity of the
central engine, using the observed H$\alpha$ luminosity of the broad line component,
$L({\rm H\alpha}) = 5.5 \times 10^{39}$ erg s$^{-1}$ at a distance of 14.5 Mpc
(Storchi-Bergmann et al. 1993), we may estimate $L_{\rm bol} \sim 100 L({\rm H\alpha})
\simeq 5.5 \times 10^{41}$ erg s$^{-1}$ (Ward et al. 1987; H. Mouri, private communication).
Thus we obtain the gas accretion rate, 
${\dot M}_{\rm acc} \simeq 1.1 \times 10^{-4} ~ M_\odot$ y$^{-1}$.
Adopting that the semi-minor radius of the CNSB ring is 650 pc,  the 
width of the ring is 5 pc (Barth et al. 1995), and
$L_{\rm CNSB} = 4.9 \times 10^7 L_\odot$  (Hummel et al. 1987), we obtain
${\dot M}_{\rm rad} \simeq 1.4 \times 10^{-9}~ M_\odot$ y$^{-1}$
for $r = 100$ pc.
Since this rate is much lower than the expected value, 
we conclude that the radiative avalanche does not work in NGC 1097.

\subsection{\it Ultraluminous Infrared Galaxies}

Our final test is applied to the ultraluminous infrared galaxies
(ULIGs; Sanders et al. 1988). Because of both their huge bolometric
luminosities of the starbursts, $L_{\rm CNSB} \sim 10^{12} L_\odot$
and the compact nature of the starbursts, $R \sim$ 50 pc (Condon et al. 1991),
the radiative avalanche would work most efficiently in the ULIGs.
Since it is likely that the ULIG nuclei are surrounded by
the CNSB regions (i.e., the geometrical dilution factor $\sim$ 1),
we may obtain ${\dot M}_{\rm rad} \simeq 0.07 ~ M_\odot$ y$^{-1}$.
It is here noted that the circumnuclear molecular gas mass in ULIGs
is typically $\sim
10^{10} M_\odot$ (Scoville et al. 1991). If this gas would be supplied to
the nuclear region within a merger timescale of $\sim 10^9$ years, 
the average gas accretion rate amounts to $\sim 10 M_\odot$ y$^{-1}$. 
In fact, many numerical simulations have shown that
major mergers between two gas-rich galaxies can supply a lot of gas
within a reasonable timescale (Mihos \& Hernquist 1994b and references therein).
Since this rate is much higher than 
that by the radiative avalanche, we consider that the dynamical accretion 
driven by the merger is the dominant fueling mechanism rather than 
the radiative avalanche.

\section{DISCUSSION: AN ALTERNATIVE STARBURST-AGN CONNECTION}

Although we have shown that the radiative-avalanche fueling may not
work in the actual AGNs, it is still worth discussing  possible
starburst-AGN connections because a non-negligible number of AGNs
have indeed the CNSBs. We therefore consider another possible connection
in this section.

Given the standard scenario for AGNs, the gas fueling is one of 
the most important physical processes for triggering AGNs 
(Shlosman, Begelman, \& Frank 1990).
As far as Seyfert nuclei are concerned\footnote{Ho, Filippenko, \& Sargent (1997)
show that bars have a negligible effect on the strength of AGNs
in their sample of over 300 spiral galaxies. Therefore, we do not consider
that the gas fueling driven by bars is important.}, possible fueling mechanisms may be
either tidal triggering by a companion galaxy (Noguchi 1988;
see for a review Barnes \& Hernquist 1992), or minor merger
with a satellite galaxy (Gaskell 1985; 
Mihos \& Hernquist 1994a; Hernquist \& Mihos 1995).
Although Seyfert galaxies tend to have their companion galaxies,
the percentage of Seyfert galaxies with physical companions is 12 percent
at most (Rafanelli, Violato, \& Baruffolo 1995 and references therein).
Further, there is no preferred kind of interaction (prograde, polar, or
retrograde) among the Seyfert galaxies with physical companions (Keel 1996)
although the efficient fueling would occur in prograde interacting systems.
Therefore, the majority of Seyfert galaxies have no relation with tidal 
interaction and thus should be triggered by  certain internal mechanisms.
On the other hand, 
since most galaxies have their satellite galaxies (Zaritsky et al. 1997
and references therein), it is likely that 
they have already experienced some minor mergers during their lives
(Ostriker \& Tremaine 1975; Tremaine 1981).
Hence it is suggested strongly that the minor merger hypothesis
has the great advantage rather than the tidal triggering.

Here we propose a new possible starburst-AGN connection based on the minor
merger hypothesis. Recently, Taniguchi \& Wada (1996) argued that a minor merger
with a {\it nucleated} satellite causes the efficient gas fueling,
leading  both to circumnuclear starbursts
 and then to nuclear starbursts, because of the
dynamical disturbance driven by
a supermassive binary (i.e., the host nucleus and the satellite
one) during the course of the minor merger.
If a host galaxy disk has abundant gas, a circumnuclear gas disk would be formed
prior to the attack by the satellite nucleus (Hernquist \& Mihos 1995).
Therefore, if this is the case, circumnuclear starburst would occur in an early
stage of the supermassive binary formation. 
As the separation between the nuclei decreases,
the gas clouds are channeled gradually into the host nucleus
during the course of merger evolution.
The merger timescale from a radius of $\sim$ 1 kpc to the nuclear region
may be of order $10^8$ years (Taniguchi \& Wada 1996).
Accordingly, we 
are able to explain the simultaneous presence of both {\it older}
CNSB regions and the fueled AGNs; i.e., Seyfert nuclei
with CNSBs (the majority may be S2s).
On the other hand, if a host galaxy disk has little gas, no CNSB would occur but 
the central engine would be fueled 
finally because any disk galaxies may have a bit of nuclear gas\footnote{Typical
Seyfert nuclei need gas of $\sim 10^6 M_\odot$
as the fuel to sustain the central engine
if the accretion rate is $\sim 0.01 M_\odot$ y$^{-1}$ and the
duration of active phase is $\sim 10^8$ years.} 
(e.g., Taniguchi et al. 1994), giving rise to S1s  with no CNSB.
Therefore, a variety of Seyfert nuclei can arise as due to different 
gaseous contents in the hosts.

It seems hard to detect direct evidence for minor mergers in some cases
because the dynamical perturbation should be smaller significantly than
that of typical galaxy interaction. 
The long timescale of the merger, $\sim 10^9$ years, may lead to
the  smearing of the relic of minor mergers.
Thus some well-evolved minor mergers may be observed as ordinary-looking
isolated galaxies.
However, it is known that minor mergers cause the kinematic
heating of host disks (Quinn, Hernquist, \& Fullagar 1993).
Such disk galaxies may be classified as S0 or amorphous galaxies
which are frequently observed in the Seyfert hosts
(Simkin, Su, \& Schwarz 1980; MacKenty 1990).
Therefore, the minor merger hypothesis can also explain the
observed diversity of the morphological properties of Seyfert hosts
(Simkin et al. 1980; Arsenault 1989;
MacKenty 1990; Moles, M\'arquez, \& P\'erez 1995).
It is also interesting to note that 
minor merger hypothesis can also be responsible for the
observed random orientation of radio jets in Seyfert nuclei
with respect to the host disk axis as discussed by Schmitt et al. (1997
and references therein).
Finally, we mention that the merger scenario is also applicable to the more luminous
starburst-AGN (i.e., ULIG-quasar) connection
provided that major mergers between or among nucleated gas-rich galaxies
are progenitors of quasars (Sanders et al. 1988).

\vspace{0.5cm}

We would like to thank Masayuki Umemura, Jun Fukue, Shin Mineshige, Keiichi Wada, 
Toru Yamada, Hideaki Mouri, and Neil Trentham for useful discussion and comments.
This work was financially supported in part by Grant-in-Aids for the Scientific
Research (No. 0704405) of the Japanese Ministry of
Education, Culture, Sport, and Science.


\end{document}